\documentclass{elsart}

\usepackage[T1]{fontenc}
\usepackage[mathscr]{eucal}
\usepackage{graphicx}
\usepackage{pstricks}
\usepackage{wrapfig}
\usepackage{amsmath}
\usepackage{amssymb}
\usepackage{remreset}

\usepackage{amssymb}

\begin{document}

\begin{frontmatter}



\title{Correlation in the velocity of a Brownian particle induced by frictional anisotropy and magnetic field }


\author{N. Voropajeva} and
\author{T. \"{O}rd\corauthref{cor}}
\corauth[cor]{corresponding author} \ead{teet.ord@ut.ee}

\address{Institute of Theoretical Physics, University
of Tartu,\\ 4 T\"{a}he Str., 51010 Tartu, Estonia}

\begin{abstract}
We study the motion of charged Brownian particles in an external magnetic field. It is found that a correlation appears between the components of particle velocity in the case of anisotropic
friction, approaching asymptotically zero in the stationary limit. If magnetic field is smaller compared to the critical value, determined by frictional anisotropy, the relaxation of the
correlation is non-oscillating in time. However, in a larger magnetic field this relaxation becomes oscillating. The phenomenon is related to the statistical dependence of the components of
transformed random force caused by the simultaneous influence of magnetic field and anisotropic dissipation.
\end{abstract}

\end{frontmatter}

\section{Introduction}

The problem of Brownian motion in an external magnetic field was first investigated in Refs. \cite{taylor, kur1, kur2} in connection with the diffusive processes in plasma. About forty years
later certain developments \cite{czo, tan, holod, zag, jim} appeared in this area again. In particular, anisotropic diffusion across an external magnetic field was considered in Refs.
\cite{holod, zag}.

In the present letter we demonstrate the appearance of the correlations in the velocity of charged Brownian particles caused by magnetic field. The effect is possible only in environment
characterized by anisotropic friction. In what follows we will use a scheme, where the deterministic parts of the stochastic equations of motion are transformed into independent equations. It
simplifies substantially the derivation of Fokker-Planck equation in velocity space as well as makes more easy the understanding of the relevant physical background. The approach is different
from the method of Ref. \cite{jim}, where rotating time-dependent basis was used to transform the Langevin equation in an external magnetic field.

\section{Probability distribution in velocity space}

The equations of motion of a Brownian particle in an external magnetic field $\overrightarrow{B}=(0,0,B_z)$ reads
\begin{equation}\label{1}
\frac{dv_i(t)}{dt}=-\sum_{j=1}^3 \lambda_{ij}v_j+\xi_i(t) \;, \quad i=1,2,3=x,y,z \, ,
\end{equation}
where
\begin{equation}\label{2}
    \lambda_{ij}=
    \left(
      \begin{array}{ccc}
        \beta_x & -\omega_z & 0 \\
        \omega_z & \beta_y & 0 \\
        0 & 0 & \beta_z \\
      \end{array}
    \right) \, ,
\end{equation}
$\omega_z=eB_z/mc$ is the cyclotron frequency and $\beta_{x,y,z}$ are the friction coefficients for a particle moving in the corresponding direction. Statistical properties of the random force
are given by the conditions
\begin{equation}\label{3}
    \langle\xi_i(t)\rangle=0\,, \quad \langle\xi_i(t)\xi_j(t')\rangle=
    a_i\delta_{ij}\delta(t-t') \,,
\end{equation}
where $a_i=2k_B T\beta_i/m$ are the components of the intensity of the Langevin source.

The system of equations (\ref{1}) can be transformed into new equations of motion
\begin{equation}\label{4}
\frac{du_i(t)}{dt}=-\Lambda_i u_i(t)+\zeta_i (t)\,,
\end{equation}
where
\begin{equation}\label{5}
    \Lambda_{1,2}=
    \frac{1}{2}\bigl(\beta_{x}+\beta_y\pm i\,\Omega\bigr), \quad \Lambda_3= \beta_{z} \,,
\end{equation}
\begin{equation}\label{6}
    \Omega=\sqrt{4\omega^2_z-(\beta_{x}-\beta_{y})^2} \,,
\end{equation}
\begin{equation}\label{7}
        \zeta_i(t)=
    \sum_{j=1}^3 \alpha_{ij}\xi_j(t) \,,
\end{equation}
by introducing, in general, complex velocities
\begin{equation}\label{8}
u_i=\sum_{j=1}^3 \alpha_{ij}v_j \, .
\end{equation}
Here the matrix of the velocity transformation (\ref{8}) can be chosen as
\begin{equation}\label{9}
\alpha_{ij}= \left(
      \begin{array}{ccc}
\dfrac{1}{\sqrt{1+|b_{1,2}|^{2}}} & \dfrac{-b_{1,2}}{\sqrt{1+|b_{1,2}|^{2}}} & 0 \\
\dfrac{-b_{1,2}}{\sqrt{1+|b_{1,2}|^{2}}} & \dfrac{1}{\sqrt{1+|b_{1,2}|^{2}}} & 0 \\
0 & 0 & 1
\end{array}
    \right)\, ,
\end{equation}
where
\begin{equation}\label{9+}
b_{1,2}=-\frac{\Lambda_{1,2}-\beta_x}{\omega_z} \, .
\end{equation}
In Eq. (\ref{9}) one has to take $b_{1}$ if $\beta_{x}>\beta_{y}$, and $b_{2}$ if $\beta_{x}<\beta_{y}$. Thereby it is guaranteed that the limit $\omega_z\rightarrow0$ leads to the unity
transformation, $\alpha_{ij}=\delta_{ij}$. In the case of isotropic friction $\beta_{x,y,z}=\beta$ the matrix (\ref{9}) reduces to the following unitary matrix
\begin{equation}\label{9a}
\alpha_{ij}= \left(
      \begin{array}{ccc}
\dfrac{1}{\sqrt{2}}  & \dfrac{\pm i |\omega_z|}{\sqrt{2}\omega_z} & 0 \\
\dfrac{\pm i |\omega_z|}{\sqrt{2}\omega_z} &  \dfrac{1}{\sqrt{2}}  & 0 \\
0 &  0 & 1
\end{array}
\right) \, ,
\end{equation}
being independent of magnetic field strength and friction coefficient $\beta$. The latter transform is in certain sense close to the approach used in Ref. \cite{lan} in the examination of the
deterministic motion of a charged particle in a magnetic field. The choice of sign in Eq. (\ref{9a}) is arbitrary.

On the basis of the Langevin equations (\ref{4}) one obtains the Fokker-Planck equation
\begin{equation}\label{11}
    \frac{dW(\vec u,t|\vec u_0)}{dt}=
    \!\!\sum_{i=1}^3 \!\frac{\partial}{\partial u_i}
    \left[\Lambda_i u_iW(\vec u,t|\vec u_0)\right]+
    \sum_{i,j=1}^3 \frac{A_{ij}}{2} \,
    \frac{\partial^2W(\vec u,t|\vec u_0)}{\partial u_i \partial
    u_j}
\end{equation}
with the initial condition $W(\vec u,0|\vec u_0)=\delta(\vec u-\vec u_0)$. In Eq. (\ref{11})
\begin{equation}\label{10}
A_{ij}=\sum_{k=1}^3 \alpha_{ik}\alpha_{jk}a_{k}\,.
\end{equation}

The solution of Eq. (\ref{11}) in terms of the velocity $\overrightarrow{v}$ is
\begin{equation}\label{12}
    W(\vec v,t|\vec v_0)=\sqrt{\frac{1}{(2\pi)^3det[h_{ij}]}}\,
    exp\left[-\frac12\sum_{i,j=1}^3
    \left(h^{-1}\right)_{ij}(v_i-\langle v_i\rangle)(v_j-\langle v_j\rangle)\right]\,,
\end{equation}
where
\begin{equation}\label{13}
    h_{ij}=\left(
      \begin{array}{ccc}
    \dfrac{K_{22}}{K_{11}K_{22}-K_{12}^{\,2}} & -\dfrac{K_{12}}{K_{11}K_{22}-K_{12}^{\,2}} & 0 \\
    -\dfrac{K_{12}}{K_{11}K_{22}-K_{12}^{\,2}} & \dfrac{K_{11}}{K_{11}K_{22}-K_{12}^{\,2}} & 0  \\
    0 & 0 & \dfrac{1}{K_{33}}
  \end{array}
    \right) \,,
\end{equation}
and the averaged values of the components of velocity are given by
\begin{eqnarray}\label{14}
   \quad \quad\quad\quad\quad\quad\quad\quad \langle v_x\rangle &=& \frac{K_2K_{12}-K_1K_{22}}{2(K_{11}K_{22}-K_{12}^{\,2})} \; ,\nonumber\\
    \langle v_y\rangle &=& \frac{K_1K_{12}-K_2K_{11}}{2(K_{11}K_{22}-K_{12}^{\,2})} \; ,\nonumber\\
    \langle v_z\rangle &=& \frac{-K_3}{2K_{33}} \; .
\end{eqnarray}
The following notations have been used in Eqs. (\ref{13}) and (\ref{14}):
\begin{eqnarray}\label{15}
    K_{ij}&=&
    \frac{1}{g}\biggl[\frac{(\alpha_{21})^{2}a_{1}+(\alpha_{22})^{2}a_{2}}{2\Lambda_2}\,\frac{\alpha_{1i}\,\alpha_{1j}}{\varphi_{11}}+
    \frac{(\alpha_{11})^2a_{1}+(\alpha_{12})^{2}a_{2}}{2\Lambda_1}\,\frac{\alpha_{2i}\,\alpha_{2j}}{\varphi_{22}}\nonumber\\
    &-\!&\frac{\alpha_{11}\alpha_{21}a_{1}+\alpha_{12}\alpha_{22}a_{2}}{\Lambda_1+\Lambda_2}\,
    (\alpha_{1i}\,\alpha_{2j}+\alpha_{1j}\,\alpha_{2i})\frac{\varphi_{12}}{\varphi_{11}\varphi_{22}}\biggr]\:; \quad i,j=1,2 \:, \nonumber\\
K_{33}&=&\frac{2\Lambda_3}{a_{3}}\,\varphi_{33}^{-1}\:,
\end{eqnarray}
\begin{eqnarray}\label{15a}
K_i&=&\frac{-2}{g}\biggl\{\Bigl[\alpha_{11}f_{11}+\alpha_{12}f_{12}\Bigr]
     \biggl[\frac{(\alpha_{21})^2a_{1}+(\alpha_{22})^{2}a_{2}}{2\Lambda_2}\,\frac{\alpha_{1i}}{\varphi_{11}}\nonumber\\
     &-&\frac{\alpha_{11}\alpha_{21}a_{1}+\alpha_{12}\alpha_{22}a_{2}}{\Lambda_1+\Lambda_2}\,
     \frac{\varphi_{12}}{\varphi_{11}\varphi_{22}}\,\alpha_{2i}\biggr]\nonumber\\
     &+&\Bigl[\alpha_{21}f_{21}+\alpha_{22}f_{22}\Bigr]
     \biggl[\frac{(\alpha_{11})^2a_{1}+(\alpha_{12})^{2}a_{2}}{2\Lambda_1}\,\frac{\alpha_{2i}}{\varphi_{22}}\nonumber\\
     &-&\frac{\alpha_{11}\alpha_{21}a_{1}+\alpha_{12}\alpha_{22}a_{2}}{\Lambda_1+\Lambda_2}\,
     \frac{\varphi_{12}}{\varphi_{11}\varphi_{22}}\,\alpha_{1i}\biggr]\biggr\} \:;  \quad i=1,2 \:, \nonumber\\
     K_3&=&-\frac{4\Lambda_3}{a_{3}}f_{33}\,\varphi_{33}^{-1}\:,
     \end{eqnarray}
\begin{eqnarray}\label{15b}
g&=&\frac{\left({\alpha_{11}}^{2}a_{1}+{\alpha_{12}}^{2}a_{2}\right)
    \left({\alpha_{21}}^{2}a_{1}+{\alpha_{22}}^{2}a_{2}\right)}{4\Lambda_1\Lambda_2}\nonumber\\
    &-&\frac{(\alpha_{11}\alpha_{21}a_{1}+\alpha_{12}\alpha_{22}a_{2})^2}{(\Lambda_1+\Lambda_2)^2}
    \frac{{\varphi_{12}}^2}{\varphi_{11}\varphi_{22}}\,,
    \end{eqnarray}
\begin{eqnarray}\label{15c}
\varphi_{ij}&=&1-e^{-(\Lambda_i+\Lambda_j)t}\:, \quad f_{ij}=v_{0j}e^{-\Lambda_it}\:.
\end{eqnarray}

\section{Correlation between the components of velocity}

Now we concentrate attention to the correlation between the components of the velocity of a particle, perpendicular to a magnetic field, $v_{x}$ and $v_{y}$. The correlation function
\begin{eqnarray}\label{16}
k(t)=\langle v_{x}(t)v_{y}(t)\rangle-\langle v_{x}(t)\rangle \langle v_{y}(t)\rangle
\end{eqnarray}
is determined by Eqs. (\ref{12}) and (\ref{13}), which yield $k(t)=h_{12}$. As a result we have on the basis of Eqs. (\ref{5}), (\ref{9}), (\ref{13}) and (\ref{15})
\begin{eqnarray}\label{17}
k(t)=\frac{-4k_B T}{m}\omega_z(\beta_x-\beta_y)\left[\frac{
    sin\left(\frac{\Omega t}{2}\right)
    }{\Omega}\right]^2\,e^{-(\beta_x+\beta_y)t}\,.
\end{eqnarray}
\begin{figure}[b]
\begin{center}
\includegraphics[width=0.6\linewidth]
{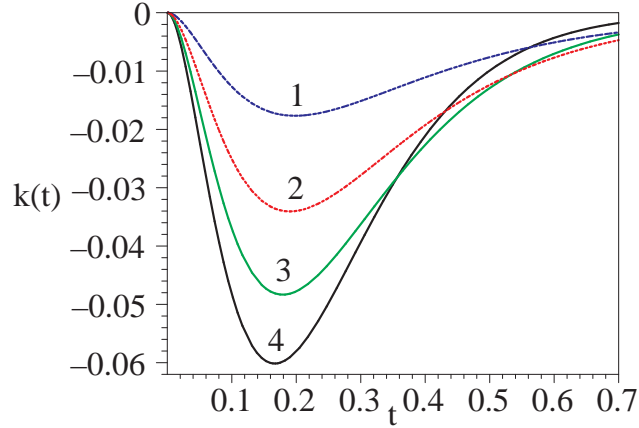} \caption{Dependence of the correlation function $k(t)$ on time at $k_B T/m=1$ and $\beta_{x}=10$, $\beta_{y}=2$ for various values of cyclotron frequency
$\omega_{z}\leq\omega_{z}^{cr}=4$. Curve 1: $\omega_{z}=1$, curve 2: $\omega_{z}=2$, curve 3: $\omega_{z}=3$, curve 4: $\omega_{z}=\omega_{z}^{cr}$.}
\end{center}
\end{figure}

Consequently, the correlation function (\ref{16}) is nonzero only if $\omega_z\neq0$ and $\beta_x\neq\beta_y$, i.e. in the presence of external magnetic field and frictional anisotropy. The
correlation approaches asymptotically zero in the stationary limit ($t\rightarrow\infty$). In this relaxation process one can distinguish two regimes. The time dependence of $k(t)$ is
non-oscillating (see Fig. 1) if $|\omega_{z}|<\omega_{z}^{cr}$. The oscillating behavior (see Fig. 2) appears if $|\omega_{z}|>\omega_{z}^{cr}$. Here
\begin{eqnarray}\label{18}
\omega_{z}^{cr}=\frac{|\beta_{x}-\beta_{y}|}{2}
\end{eqnarray}
is the critical value of the cyclotron frequency separating the regions where $\Omega$ in Eq. (\ref{17}) is imaginary or real quantity correspondingly.

\begin{figure}[t]
\begin{center}
\includegraphics[width=0.6\linewidth]
{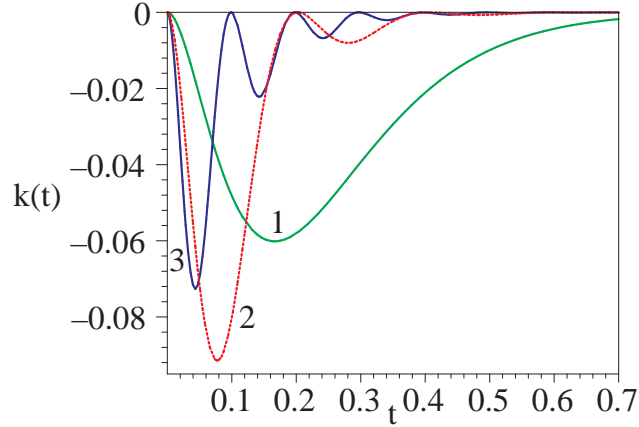} \caption{Dependence of the correlation function $k(t)$ on time at $k_B T/m=1$ and $\beta_{x}=10$, $\beta_{y}=2$ for various values of cyclotron frequency
$\omega_{z}\geq\omega_{z}^{cr}=4$. Curve 1: $\omega_{z}=\omega_{z}^{cr}$, curve 2: $\omega_{z}=16$, curve 3: $\omega_{z}=32$.}
\end{center}
\end{figure}

\section{Discussion}

In conclusion, we have found in the short time-scale the correlation between the components of the velocity of a charged Brownian particle caused by external magnetic field and frictional
anisotropy. The effect arises\footnote{Note also, that the necessary condition for the appearance of the effect is non-zero temperature (see Eq. (\ref{17})), which unambiguously indicates to the
entirely stochastic nature of the phenomenon.} due to the statistical dependence of the components of transformed random force $\zeta_{1,2}(t)$, determined by Eq. (\ref{7}).  On the basis of
Eqs. (\ref{7}) and (\ref{3}) we have
\begin{equation}\label{19}
\langle\zeta_1(t)\zeta^{*}_2(t')\rangle=\sum_{k} \alpha_{1k}\alpha^{*}_{2k} a_k \delta(t-t')\, .
\end{equation}
Whereas in the general case of anisotropic friction and non-zero magnetic field the sum $\sum_{k}\alpha_{1k}\alpha^{*}_{2k}a_k$ is not equal to zero, it is impossible to transform the system
(\ref{1}) into entirely independent equations. Although the deterministic part of the equations of motion (\ref{1}) can be decoupled, the channel of correlation appears in Eqs. (\ref{4}) for the
components of stochastic force in presence of an external magnetic field and anisotropic dissipation. As a result the components of the velocity of a Brownian particle, perpendicular to the
magnetic field, turn out to be correlated.

However, in the case of isotropic friction the expression
\begin{equation}\label{20}
\langle\zeta_i(t)\zeta^{*}_j(t')\rangle=\frac{2k_B T\beta}{m}\delta_{ij}\delta(t-t')
\end{equation}
is valid for the arbitrary components of random force $\zeta_i(t)$ because the condition of unitary transformation, $\sum_{k}\alpha_{ik}\alpha^{*}_{jk}=\delta_{ij}$, holds. In this situation the
system of equations (\ref{1}) can be entirely decoupled and the correlation between the components of velocity is absent. We obtain the same result for anisotropic friction if magnetic field
equals to zero, due to $\alpha_{ij}=\delta_{ij}$ in this case.

\begin{ack}
\vspace{-15pt} The authors acknowledge support by Estonian Science Foundation through Grant No. 6789.
\end{ack}



\end{document}